\pdfoutput=1
\documentclass{cdclass}
\newcommand{\detail}[1]{{\color{blue} \small #1}}
\usepackage{algorithm}

\usepackage{hyperref}
\hypersetup{
    colorlinks=true,
    }

\begin{document}

\title{Viral transmission in pedestrian crowds: Coupling an open-source code assessing the risks of airborne contagion with diverse pedestrian dynamics models}
\titlerunning{Viral transmission in pedestrian crowds} 

\author{
  Alexandre NICOLAS\authorlabel{1} \and 
  Simon MENDEZ\authorlabel{2} 
}
\authorrunning{A. Nicolas \and S. Mendez}
\institute{
  \authorlabel{1}  Institut Lumi\`ere Mati\`ere, CNRS, Univ. Lyon 1, Villeurbanne, F-69622, France 
  \authoremail{1}{alexandre.nicolas@cnrs.fr}
  \and
  \authorlabel{2}  IMAG, Univ. Montpellier, CNRS, Montpellier, F-34095, France
  \authoremail{2}{simon.mendez@umontpellier.fr}
}

\date{2023}{date1}{date2}{date3} 
\ldoi{10.17815/CD.20XX.X} 
\volume{V}  
\online{AX} 

\maketitle

\begin{abstract}
We study viral transmission in crowds via the short-ranged airborne pathway using a purely model-based approach. 
Our goal is two-pronged. Firstly, we illustrate with a concrete and pedagogical case study how to estimate the risks of new viral infections by coupling pedestrian simulations with the transmission algorithm that we recently released as open-source code. The algorithm hinges on pre-computed viral concentration maps derived from computational fluid dynamics (CFD) simulations. Secondly, we investigate to what extent the transmission risk predictions depend on the pedestrian dynamics model in use. For the simple bidirectional flow under consideration, the predictions are found to be surprisingly stable across initial conditions and models, despite the different microscopic arrangements of the simulated crowd, as long as the crowd evolves in a qualitatively similarly way. On the other hand, when major changes are observed in the crowd's behaviour, notably whenever a jam occurs at the centre of the channel, the estimated risks surge drastically.
\end{abstract}



\section{Introduction}

Designing robust tools to assess the risks of viral spread in pedestrian crowds and making them readily accessible to practitioners with limited computing capabilities is of manifest practical relevance for public safety, especially in the aftermath of the COVID-19 pandemic. For example, it would enable them to test variations of designs or flow rules for a given venue and determine which is the most suitable if an epidemic is lurking.
While a number of such tools have been proposed, they generally either rely on a very coarse, and questionable, evaluation of transmission risks \cite{harweg2021agent} or require considerable computing capabilities, even in their streamlined version \cite{lohner2021high}. In a recent publication \cite{mendez2023microscopic}, we focused on the short-ranged airborne transmission route \cite{wang2021short}, i.e., the infection of a susceptible individual after the inhalation of virus-laden aerosols emitted by a contagious person who talks, shouts, eats, or just breathes in their immediate surrounding, and we showed how to 
overcome the foregoing limitations: this was achieved by computing spatio-temporal maps of viral concentration around a contagious person using high-fidelity microscopic simulations in a variety of micro-environments and storing them in memory, so that at run time they are just loaded to assess the transmission risks between individuals in a macroscopic crowd \cite{mendez2023microscopic}. Since then, the associated Python scripts have been released as open-source code.

In this contribution, we explain how this code can be used in practice to integrate the output of crowd simulation software. Then, we investigate the sensitivity of the predictions
to the pedestrian dynamics model in use. For that purpose we consider a simple bidirectional flow and simulate it with various popular models.

\section{Principles of the model-based assessment of viral transmission risks in crowds}
\label{sec:principles}

\begin{figure}
\begin{centering}
\includegraphics[width=\textwidth]{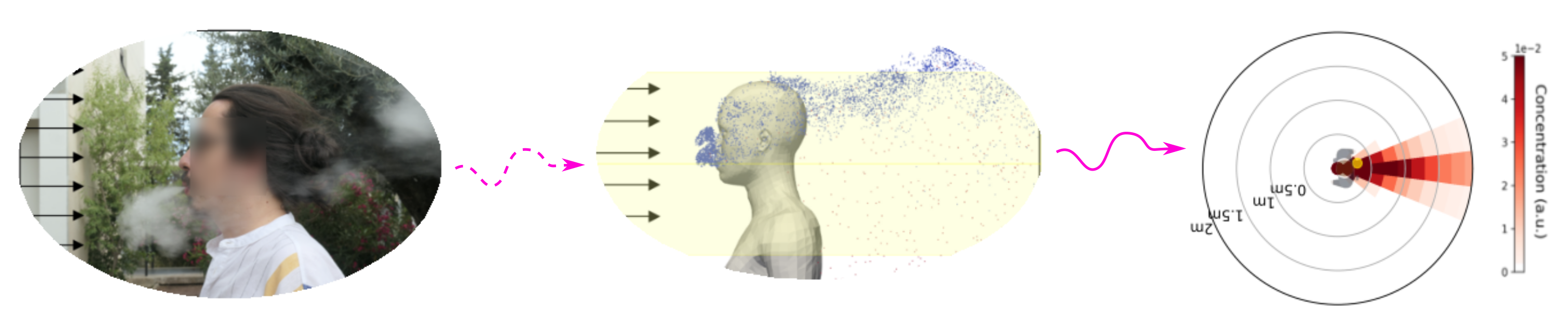}
\par\end{centering}
\caption{\label{fig:pipeline}Principle of the algorithm to assess transmission risks: Exhalations and aerosol emission are reproduced with high-fidelity CFD methods (\emph{middle}) and then coarse-grained into dynamic viral concentration maps (\emph{right}). For computational acceleration, transmission risks beyond 4~m may be discarded.}
\end{figure}

A chain of arguments is only as strong as its weakest link. Accordingly, one should not lend more credence to numerical tools designed to assess the risks of viral contagion than the underlying transmission model between individuals deserves. Consider two individuals, one contagious ($E$) and one susceptible ($R$);
the instantaneous transmission rate due to aerosols emitted by $E$ at $t_e$ and inhaled by $R$ at $t_r>t_e$ writes
\begin{equation}
\nu(t_e,t_r)=T_0^{-1}\,\tilde{\nu}\Big[E(t_e), R(t_r),\,t_r-t_e,\,\mathrm{ambient\ flows},\,\mathrm{activity(t_e)}\Big],
\label{eq:nu_def}
\end{equation}
where the function $\tilde{\nu}$, accounting for the fluid dynamics of particle emission and transport, depends on the geometric positions and head orientations of $E$ at $t_E$ and $R$ at $t_R$, the delay $t_r-t_e$, the respiratory activity of the emitter (mouth-breathing, speaking, shouting, singing, etc.), and the ambient air flows. Importantly, all pathological and physiological uncertainties and unknowns about the emitter's viral load, the
inhalation probability and the minimal infectious dose are subsumed into the variable $T_0$, which is
the characteristic time for infection in a reference situation. (In our simulations, we systematically find that the estimated transmission rates, multiplied by $T_0$, are hardly sensitive to the value of $T_{0}$.)
Early risk assessment endeavours coarsely assumed that the transmission function $\tilde{\nu}$ only depends on the distance between $E$ and $R$ at emission time $t_E$ and, quite often, that $\tilde{\nu}$ is finite and constant below a threshold distance called contagion radius, and vanishes above \cite{harweg2021agent}. This is of course very questionable and we will see in Sec.~\ref{sec:Impact_ped_model} that this approach may be misguided.
At the other extreme, detailed CFD simulations take into account all geometric considerations, the spatial layout, the air flow field and other environmental details; they are thus very accurate, but sensitive to small variations in the input conditions and, above all, computationally very demanding \cite{abkarian2020speech,lohner2021high}.

To limit this computational cost, we adopted an intermediate stance in \cite{mendez2023microscopic}, by
simulating inhalation and exhalation flows beforehand in a variety of micro-environmental conditions, notably for a gamut of relative winds, and injecting in the resulting flow aerosols and droplets with a size distribution similar to that found while either mouth-breathing or talking. From the simulated microscopic trajectories of these particles, we derived coarse-grained spatio-temporal maps of viral concentration around a contagious emitter taken in isolation and stored them in memory; the workflow is sketched in Fig.~\ref{fig:pipeline}. Risks of new infections can then be obtained very efficiently by coupling these maps to pedestrian trajectory and head orientation data.

\begin{figure}
\begin{centering}
\includegraphics[width=0.7\textwidth]{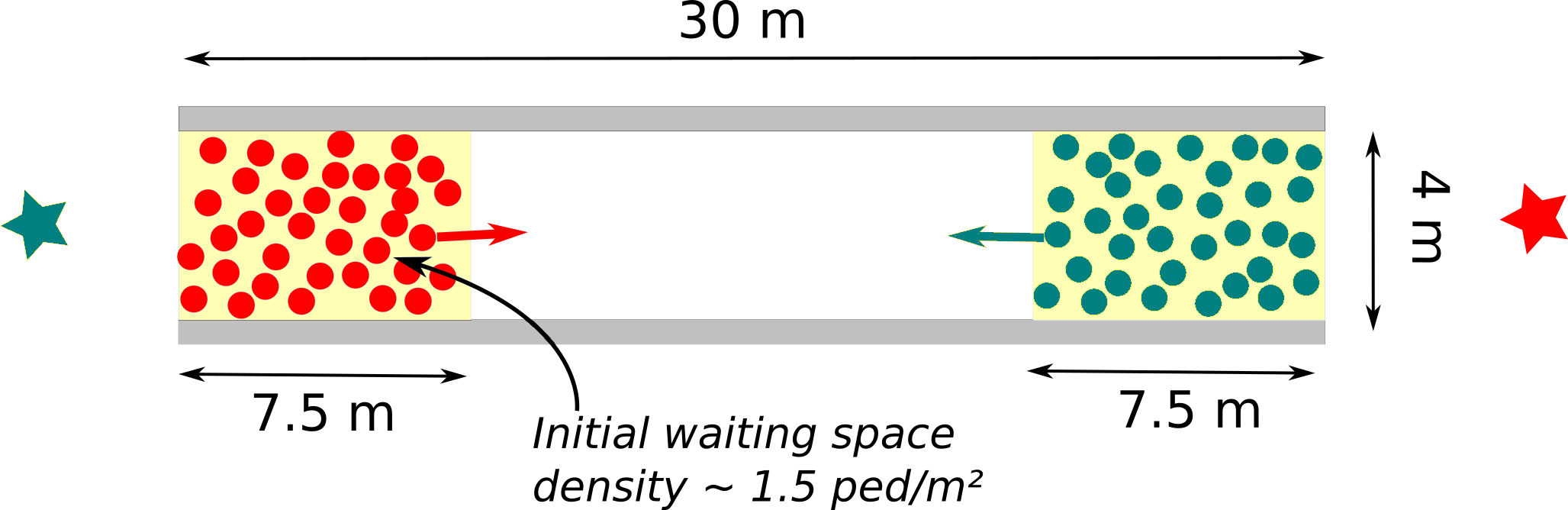}
\par\end{centering}
\caption{\label{fig:sketch}Sketch of the geometry and initial conditions of the considered bidirectional corridor flow configuration. 45 blue agents head for the blue star to the left, whereas their 45 red counterparts head right.}
\end{figure}

While these data were obtained from field observations in our previous works \cite{garcia2021model,mendez2023microscopic}, here we explore the possibility of using numerical simulations of crowd dynamics as input to the risk assessment code. To gauge how likely the results are to be robust, we shall use widely different pedestrian models. In terms of scenario, particular simple flow settings are used, namely, a well-balanced bidirectional flow in a 4-meter-wide
corridor, which is located either outdoors or in a well ventilated space (so that risks
mostly originate from direct exposure to emitted droplets and aerosols,
rather than airborne contagion via long suspended aerosols). The layout and initial conditions are shown in Fig.~\ref{fig:sketch}: initially, the agents (of body radius 0.2 m) are randomly positioned in their respective
waiting spaces, without overlap; their preferential
speeds are randomly drawn from a uniform distribution between 1.0 and 1.5~m/s.
Let us underline that this flow scenario is devoid of many
complexities which would add to the modeller's burden in many practical situations, such as the 
uncertainty about people's destinations, their possible wish to halt temporarily, or the presence of social groups.

\section{Our open-source code -- a succinct tutorial}

Suppose that the pedestrian trajectories and head orientations are already known (this will be  at the heart of the next section). To estimate the rate of new infections caused by an index patient,
we use the open-source Python scripts available in the public GitHub repository
\url{https://github.com/an363/InfectiousRisksAcrossScales}. After cloning
the GitHub repository and unzipping the compressed folders (notably \emph{Diagrams.zip}),
we open the input file called "\emph{InputFile.txt}" to adjust the paths and simulation conditions, viz.,
\begin{algorithm}
DiagramsFolder= \detail{[PATH TO FOLDER WITH SPATIO-TEMPORAL DIAGRAMS]}\\
OutputFolder= \detail{[PATH TO OUTPUT FOLDER]}\\
TrajectoryFolder=\detail{[PATH TO FOLDER WITH TRAJECTORIES]}\\
T0= \detail{[\emph{characteristic infection time}, in seconds]}\\
(vx,vy)= (\detail{$v_x$},\ \detail{$v_y$ in m/s}) \# external wind speed\\
ExhalationMode= \detail{[choose between: \emph{breathing / speaking / large\_droplets}]}\\
IsotropicInhalation= \detail{[choose between: \emph{True / False}]}\\
ContagionAmidGroups=\detail{[choose between: \emph{True / False}]}
\end{algorithm}

The algorithm will thus load the trajectories contained in text files within \emph{TrajectoryFolder}. Each file should correspond to one agent and should be named \detail{[GROUP-ID]}\_\detail{[PED-ID].csv}, e.g., 1\_3.csv, for pedestrian 3 belonging to social group 1; each line of this file refers to a successive timestamp and contains four floats, as follows:
\begin{algorithm}
\detail{time} (in seconds); \detail{x-position} in meters; \detail{y-position} in meters; \detail{head orientation} in radian
\end{algorithm}

For instance, "10.1;1.2;2.3;0.1" means that at time $t=10.1\,\mathrm{s}$ the agents stands at $x=1.2\,\mathrm{m}$, $y=2.3\,\mathrm{m}$, with a head oriented along a direction 0.1 rad past the $x$-axis (counter-clockwise).
Finally, we run the script \emph{main.py} with Python 3 (for instance by typing \$ 
\detail{python3 main.py} in the terminal).

The \emph{characteristic infection time} is the timescale $T_0$ appearing in Eq.~\ref{eq:nu_def}; 
The \emph{ExhalationMode} determines the number and size distribution of emitted droplets;
the Booleans \emph{IsotropicInhalation} and \emph{ContagionAmidGroups} indicate if agents can inhale aerosols coming hitting the back of their heads (by default, we recommend to set it to \emph{False}) and if a contagious person can infect other members of their social group (i.e., if these members are still
susceptible despite probable previous contacts with the index patient).

Note that semi-colons can be used to separate multiple input conditions if the user wants to launch multiple runs sequentially with a single input file, for instance \detail{(vx,vy)=(0.0,0.0);} \detail{(-0.2,0.5)} or \detail{ExhalationMode=speaking;breathing}. Comments after a hash symbol at the end of each line are discarded.

The risk assessment outcome is saved in \emph{OutputFolder},
with a distinct folder for each set of conditions. The folder contains a summary of the parameters (\emph{parameters.txt}), 
a file detailing how many new cases each distinct agent would cause per hour, should they be contagious (\emph{Risks\_by\_person\_output...dat}), and one file containing the mean number of new cases per hour (\emph{Risks\_mean\_output...dat}). In these two files, for every line, a lower bound \emph{Clow} and an upper bound \emph{Cbar} on the number of new cases are given: the upper bound is the value if everybody (but the index patient)
is susceptible at the beginning of the simulation, whereas the lower bound refers to the possibility that agents have already been infected by the index patient, outside the observed field of view. In practice, these two bounds are extremely close to one another, so we shall not elaborate more in this regard.

\section{Impact of the chosen pedestrian model on the risk assessment}
\label{sec:Impact_ped_model}
To determine to what extent the risks of new infections in a given setup can be determined by a fully model-based approach, 
we now couple the foregoing algorithm with the output of pedestrian simulation software. To this end, we consider 
the simple bidirectional-flow settings introduced in Section \ref{sec:principles} and compare the trajectory predictions of
various crowd dynamics algorithms, using the UMANS software developed at INRIA \cite{van2020generalized} to simulate them all.

\begin{figure}
\begin{centering}
\includegraphics[width=\textwidth]{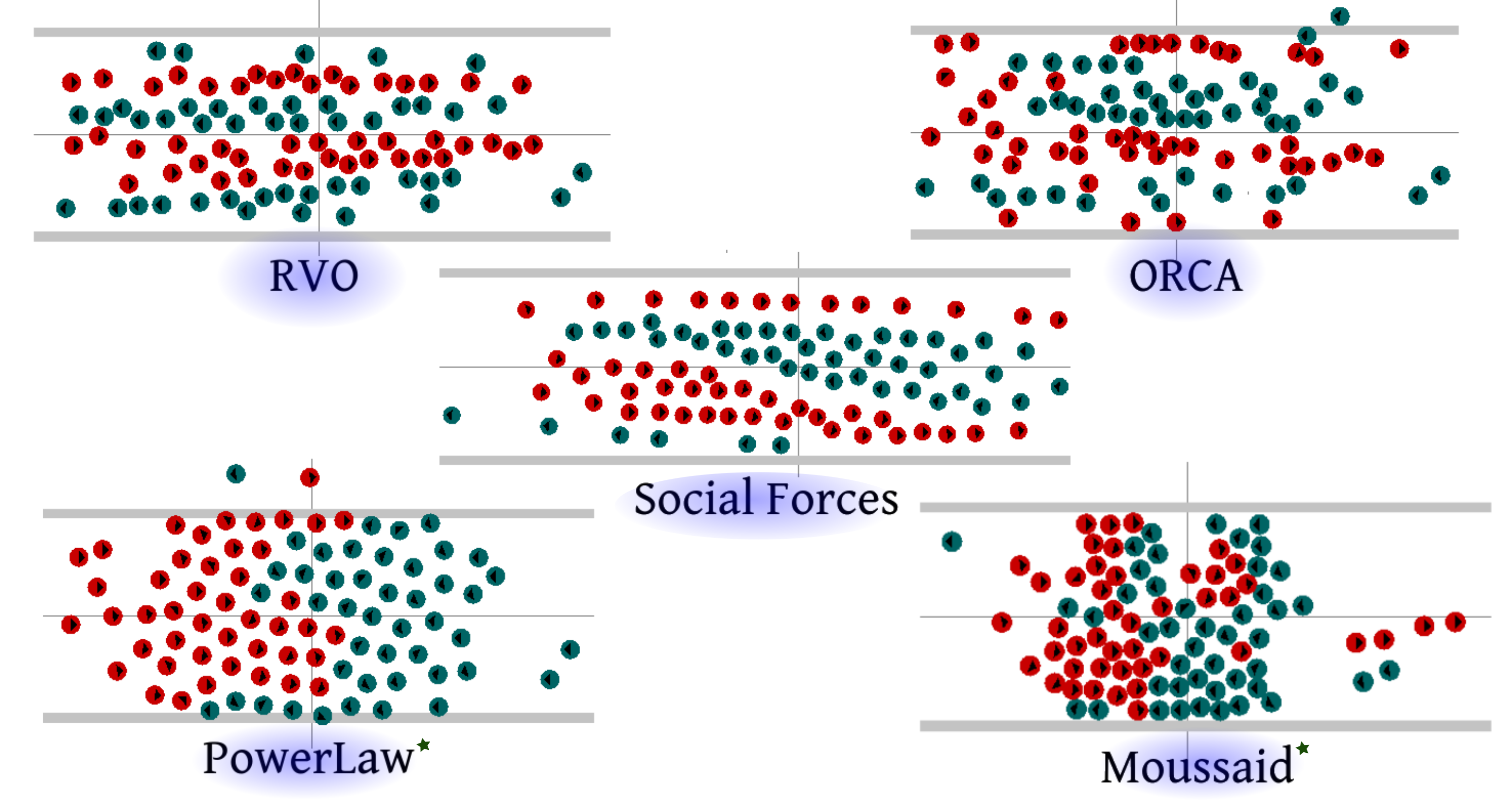}
\par\end{centering}
\caption{\label{fig:sim_snapshots}Snapshots of the simulations, approximately halfway through the run, simulated and visualised with the UMANS software, using the native parameters for every model and a constant integration time step $dt=0.01\,\mathrm{s}$ (which departs from the continuous limit at least for \emph{PowerLaw} and \emph{Moussaid}).}

\end{figure}

More precisely, we put to the probe five different models. Two come from the field of robotics, RVO \cite{van2008reciprocal}
and ORCA \cite{van2011reciprocal}, and are based on the idea of velocity obstacles: roughly speaking, each agent selects the velocity closest to its desire, but outside the `cone' of  velocities leading to an imminent collision. Instead, \emph{SocialForces} posits that pedestrians maintain their
distance to their neighbours because of binary repulsive pseudo-forces at play in a Newton-like equation \cite{helbing1995social} (these forces are sensitive to the relative velocities of the agents in the UMANS implementation). Karamouzas et al.'s \emph{PowerLaw} model \cite{Karamouzas2014universal} substitutes these mostly distance-dependent pseudo-forces with pseudo-forces that are inversely related to the anticipated time to collision (i.e., the delay after which a collision is expected if all velocities are conserved). Finally, \emph{Moussaid} 
resorts to simple heuristics to select a direction of motion (which should bring the agent closest to the goal before any collision) and a speed \cite{moussaid2011simple}. In all simulations, pedestrians are described as disks of radius 0.2 m and preferential speeds are uniformly distributed between 1 and 1.5 m/s.  The predictions of the models at a random time in the scenario are shown in Fig.~\ref{fig:sim_snapshots}

Importantly, we do not optimise the parameters of the models,
but use the default ones natively coded in the UMANS software. Furthermore, we adopt a constant numerical time step $dt=0.01\,\mathrm{s}$ for all models.
Clearly, should one wish to make the simulation as realistic as possible, then  the parameters and time step
would need to be adjusted. In particular, $dt=0.01\,\mathrm{s}$ is definitely too large
for some simulations, which leads to occasional spurious effects (such as a reduced ability to form lanes in \emph{Moussaid} and \emph{PowerLaw}). Last but not least, the corridor walls are not properly handled in the UMANS implementation of the \emph{PowerLaw} and \emph{ORCA} models, so that a couple of agents cross them now and them (but most agents remain confined in the corridor width).
While these artifacts are undesirable for practical applications,
here they will actually turn out to be quite enlightening, as we shall see, because they give rise to a different phenomenology whose impact on our risk assessments we can explore.

Let us now couple the simulated pedestrian trajectories with the viral transmission code. To do so, we consider environmental conditions with very little wind (actually, an almost undetectable draught), $(v_x,v_y)=(0,0.2\,\mathrm{m/s})$, $T_0=900\,\mathrm{s}$ and we assume that people are constantly talking in the scenario, with their heads oriented in the instantaneous walking direction and an emission point at the centre of the head. 

\begin{figure}
\begin{centering}
\includegraphics[width=0.7\textwidth]{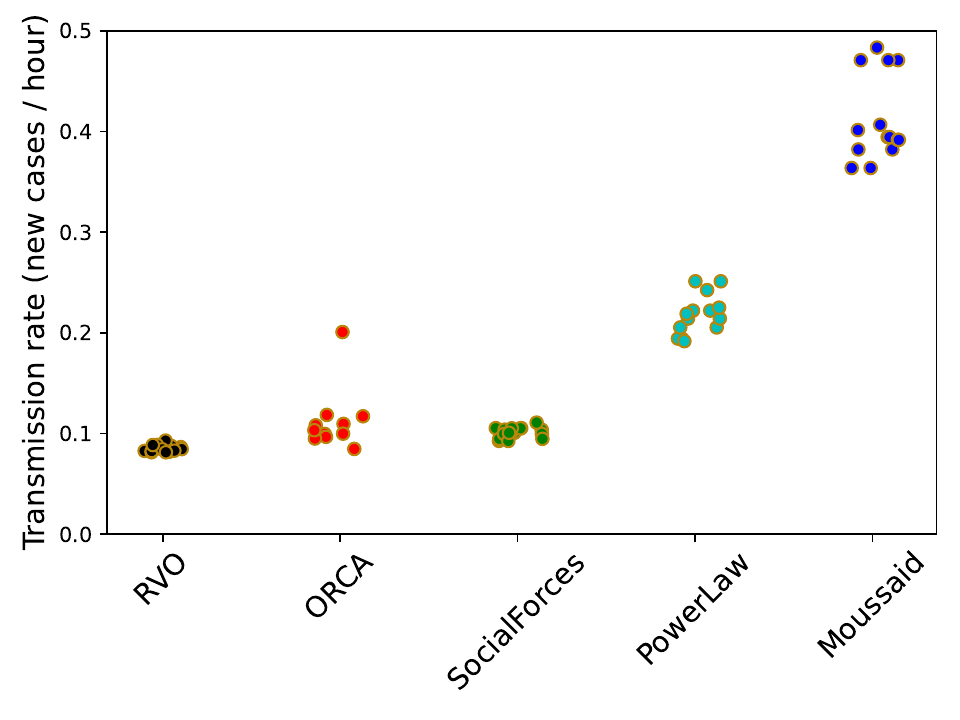}
\par\end{centering}
\caption{\label{fig:results}Mean umber of new cases per hour caused by a contagious agent, estimated using diverse pedestrian simulation models and varying the initial conditions.}
\end{figure}

In these conditions, Fig.~\ref{fig:results} presents the estimated transmission rates for at least a dozen simulations of each model for 18 seconds each (each realisation is represented by a dot).
We start by focusing on the three leftmost columns, corresponding to \emph{RVO}, \emph{ORCA}, and \emph{Social Forces}.
What strikes us is the strong homogeneity (low dispersion) within each model, i.e., across all realisations of initial positions  and preferential speeds. These variations visibly affect the dynamics, but they hardly impact the estimated transmission rates, apart from one outlier in the ORCA model (to which we shall come back later). 
Even more arresting is the consistency of the predictions across the three models. This is partly
unexpected because the arrangement of the crowd, notably in terms of spacings, differs markedly between the models, as illustrated in Fig.~\ref{fig:sim_snapshots}. In particular, \emph{ORCA} agents tend to come very close to each other, almost within contact, which is consistent with the fact that velocity obstacles prohibit the selection of a velocity that leads to a collision, but do not penalise close proximity. By contrast, \emph{SocialForces} agents, repelled by their mutual social forces, keep a significant distance to each other. This discrepancy would have led to drastic variations in the estimates if we had used a na\"ive transmission model based on the instantaneous spacing between agents, or on a distance-based contagion radius. Instead, here, because pedestrians are moving, the virus-rich exhalation cloud dragged in their wake may affect people following them;  the external wind and the walking-induced relative wind also undermine any na\"ive assumption of an isotropic decay of the transmission risks with distance.

Bearing in mind this strong homogeneity in the risks estimates despite the observed differences in the dynamics, how can one explain the much larger estimates obtained with the \emph{PowerLaw} and, even more so, \emph{Moussaid} models, as well as for one \emph{ORCA} outlier? Direct visualisation of the numerical output gives the answer: In these simulations (including the \emph{ORCA} outlier), the two counter-flows fail to intertwine so as to pass each other and, instead, form a jam at the centre of the channel; these jams are pretty systematic, but may be temporary (as with \emph{PowerLaw}) or much longer-lived (\emph{Moussaid}). In that situation, agents in the jam stand close to each other, at almost zero speed. In the absence of significant external wind to disperse the exhaled puffs, this configuration is quite favorable for airborne transmission, which explains the high transmission rates.

\section{Conclusion}

In summary, we have explored the possibility of a purely model-based assessment of the risks of airborne viral transmission (by direct exposure), in actual and hypothetical scenarios.
Using a practical example, we have explained
how to make use of our recently released Python scripts designed for that purpose. The method hinges on the assumption that an agent's puff is only marginally influenced by the other pedestrians in the neighbourhood and relies on pre-computed dynamic viral concentration maps, which are computed with detailed CFD simulations in a variety of micro-environmental conditions and stored in memory. This leads to robust model-based risk assessments that are several orders of magnitude faster than CFD-based ones and can be run on a personal laptop.

Then, enquiring into the role of the pedestrian model in use to simulate pedestrian trajectories, we have compared the output of diverse models implemented in the UMANS software \cite{van2020generalized}, in a non-specifically tailored version. This has led to interesting, partly unexpected findings. In the very simple \emph{dynamic} settings under study, the fine-grained organisation of the crowd flow has little incidence on the global transmission rate. In particular, the estimated rates are remarkably stable with
respect to the initial positions and preferential speeds of the agents, and even to how much social distance they maintain around themselves (e.g., very little in \emph{ORCA}, much more in \emph{SocialForces}). This is quite at odds with what would be predicted by models premised on the idea of a contagion radius; it can be rationalised by the aerosols left and dragged in pedestrians' wakes (noting the absence of any prolonged tight contact in the present scenario). On the other hand,  substantial variations in transmission risks \emph{do} occur if major phenomenological changes take place in the crowd. For instance, a drastic increase in transmission rates is observed whenever the counter-walking groups fail to form intercalating lanes that fluidise the flow, but end up forming a high-density, halted jam at the centre. More broadly speaking, we expect that, for an identical scenario with \emph{moving} people, moderate variations in the trajectories predicted by distinct models will have limited incidence on viral spread estimates, whereas structural differences (such as the formation of groups of halted people, possibly talking to each other) will substantially alter the results; these expectations may not hold for \emph{static} scenarios in windless conditions, where the transmission risks may sensitively depend on the distance between people.

\paragraph*{Author contributions.}
Conceptualisation, methods: both authors. CFD simulations: SM. Python scripts: AN. First draught: AN.

\paragraph*{Acknowledgements.}
Part of this work was funded by Agence Nationale de la Recherche: projects SeparationsPietons (ANR-20-COV1-0003, A. Nicolas) and TransporTable (ANR-21-CO15-0002, S. Mendez).

\paragraph*{Conflict of interest.}
The authors are not aware of any conflict of interests.

\end{document}